\documentstyle[twocolumn,prl,aps]{revtex}

\begin{document}

\renewcommand{\topfraction}{0.8} 
\twocolumn[\hsize\textwidth\columnwidth\hsize\csname
@twocolumnfalse\endcsname
\title{On properties of elementary excitations in fractal media}
\author{A.A. Kirillov}
\address{Institute for Applied Mathematics and Cybernetics 10 Uljanova Str., Nizhny\\
Novgorod, 603005, Russia\\
e-mail: kirillov@unn.ac.ru}
\date{\today} \maketitle

\begin{abstract}
It is shown that elementary excitations in fractal media obey the so-called
parastatistics of a variable order. We show that the order of the
parastatistics $N( k) $ is a function of wave numbers $k$ which
depends on the fractal dimension $D$ as $N( k) \sim k^{D-3}$ and
represents a specific characteristic of such media. This function $N(
k) $ defines properties of the ground state for excitations and the
behavior of the spectrum of thermal fluctuations. In particular, in fractal
media fluctuations of the density acquire an amplification 
by the factor $N( \omega ) \sim \omega ^{D-3 }$ which can be related
to the origin of $1/f$-noise.
\end{abstract}

 \vskip2pc]


It is well known that a physical system in which particle distribution
follows, in some range of scales, a fractal law $N\left( r\right) \sim r^{D}$
(where $N\left( r\right) $ is the number of particles contained within a
radius $r$) with $D$ $<3$ is not an exotic example, rather it represents a
typical situation (e.g., see Refs. \cite{Fr} -\cite{Fr3} and references
therein). Nevertheless, an acceptable quantitative theory which allows for
the description of thermodynamical properties of such systems is still
absent. In the present paper we, however, show that the possibility to
construct such a theory has recently appeared. First, we note that
thermodynamical properties of a system are determined by properties and
behaviour of elementary excitations in the system. In quantum theory
excitations are described in terms of respective particles, while the number
of degrees of freedom of a particular excitation can be characterized, in
the Fourier space, by the spectral number of modes in the interval of wave
numbers between $k$ and $k+dk$ as 
\begin{equation}
\nu \left( k\right) \frac{dk}{k}.  \label{m}
\end{equation}
The behaviour of the function $\nu \left( k\right) $ depends on the
dimension of the system as $\nu \left( k\right) \sim k^{D}$ and for regular
(or homogeneous) systems the exponent $D=n$ takes integral values $n=1,2,$
or $3$. In the case of fractals, however, the dimension $D$ is not an
integral number, while in a more general case (e.g., in the case of
multifractals) this function may change its behaviour for different
intervals of scales, i.e., $D=D\left( k\right) $. Moreover, a change in the
thermodynamical state of a system may lead to essential transformations in
the behaviour of the function $\nu \left( k\right) $ (e.g., in the case of
percolation systems \cite{Fr2}). Thus, an adequate description of such
systems requires that the spectral number of modes $\nu \left( k\right) $
has to be a variable which, in general, depends on wave numbers and, in the
presence of self-organization processes, on time.

In the standard field theory the spectral number of modes is always fixed by
the dimension of space. However, there exists a modification of the field
theory (MOFT) in which the number of fields (and, therefore, of field modes)
can vary. Such a modification was proposed in Ref. \cite{K99} to account for
spacetime foam effects and it was recently developed in Refs. \cite{K03}. In
particular, MOFT was shown to predict a significant amplification of the
gravitational force at galaxy scales (which is really observed and
represents the so-called dark matter problem) and a fractal law for the
galaxy distribution with dimension $D\approx 2$ \cite{K03} which is in a
full agreement with the observed picture of our Universe \cite{LMP}. Thus,
we can expect that the theoretical scheme of MOFT provides a natural basis
to describe excitations in systems with the fractal behaviour.

The measure (\ref{m}) can be used to define the number of field modes $%
N\left( k\right) $ as $\nu \left( k\right) =k^{3}N\left( k\right) /\left(
4\pi ^{2}\right) $ (to be specific we consider a fractal volume). In the
standard field theory $N\left( k\right) =1$, while in a general case $%
N\left( k\right) $ is a variable which depends on the wave number $k$ and
time (in particular, the fractal law gives $N\left( k\right) \sim k^{D-3}$).
From the mathematical standpoint this means that excitations obey the
so-called parastatistics which generalizes the standard Bose and/or Fermi
statistics and the number of field modes $N\left( k\right) $ plays the role
of the order of the parastatistics. We note that the first attempt to
generalize the Bose and Fermi statistics was made by Gentile \cite{Gentile},
while the parastatistics was suggested by Green in Ref. \cite{G53}. Since
then, the parastatistics has been studied in many papers (e.g., see Refs. 
\cite{Vol} - \cite{gov}). In particular, experimental observation of the
fractional quantum Hall effect \cite{Stromer} stimulated the interest in
various nontraditional statistics \cite{LMW,HLCS} and an attempt was made to
relate the parastatistics to high temperature superconductivity \cite{W}.
However, the parastatistics is still considered as an exotic possibility and
has not received an adequate attention (e.g., see Ref. \cite{conf}). MOFT 
\cite{K03} gives a specific realization of the parafield theory in which the
order of the parastatistics represents an additional variable. The
discussion above shows that the parastatistics of a variable order has a
straightforward relation to the formation of properties of elementary
excitations in fractal media.

Consider as an example sound waves in a system of identical $q-$particles
whose distribution in space has, at low temperatures $T\rightarrow 0$, an
irregular (i.e., fractal) character $\rho _{0}\left( x\right) $. We note
that the case of fermionic excitations can be considered in the same way.
Sound waves represent oscillations in the density $\delta \rho (x,t)=$ $\rho
\left( x,t\right) -\rho _{0}\left( x\right) $ and the velocity $\delta
u\left( x,t\right) =u\left( x,t\right) -u_{0}\left( x\right) $ of particles
(where $\rho _{0}$, $u_{0}$ are mean values which, in fractal media, have a
specific dependence on coordinates). In what follows we denote these
functions as a field $A\left( x,t\right) =(\delta \rho \left( x,t\right)
,\delta u\left( x,t\right) )$. This field can be expanded in modes 
\begin{equation}
A\left( x,t\right) =\frac{1}{\sqrt{V}}\sum_{k,\alpha }a_{k,\alpha }e_{\alpha
}\left( k\right) e^{ikx}+a_{k,\alpha }^{+}e_{\alpha }^{\ast }\left( k\right)
e^{-ikx},  \label{a}
\end{equation}
where $V$ is the volume and $e_{\alpha }\left( k\right) $ is the
polarization vector (we assume that any additional normalization factor is
included in $e_{\alpha }\left( k\right) $). The coefficients $a_{k,\alpha
}^{+}$ and $a_{k,\alpha }$ play the role of creation and annihilation
operators for elementary excitations (phonons). Thus, in the standard
picture these operators obey the relations 
\begin{equation}
a_{k,\alpha }a_{k^{\prime },\beta }^{+}-a_{k^{\prime },\beta
}^{+}a_{k,\alpha }=\delta _{k,k^{\prime }}\delta _{\alpha ,\beta },
\label{cr}
\end{equation}
which mean that phonons represent Bose particles, while the Hamiltonian for
free particles is given by 
\begin{equation}
H_{0}=\sum_{k,\alpha }\omega _{k,\alpha }a_{k,\alpha }^{+}a_{k,\alpha }\;,
\end{equation}
where $\omega _{k,\alpha }$ is the energy of a phonon.

As it was discussed above in fractal media oscillations have a more complex
character than in homogeneous media (e.g., like in ideal crystals or gases)
and for an adequate description of oscillations a single field $A\left(
x,t\right) $ is not enough. Indeed, to illustrate this statement we can
imagine the situation when the system of particles splits into a set of
almost independent thin surfaces (e.g., the system represents a fractal with
dimension $D\simeq 2)$, so that excitations for every surface can be
described by its own field $A_{i}\left( x,t\right) $ (where $i$ numerates
the surfaces). Such surfaces can have a chaotic distribution in space and,
therefore, to distinguish them in an explicit way is not so trivial. In real
fractal media such a splitting works in the Fourier space (e.g., see Ref. 
\cite{K03}) and, therefore, in a general situation we can only state that
particles of a system can be divided in groups $n=\left\{
n_{1},n_{2},...n_{N}\right\} $ ($\sum_{1}^{N}n_{i}=n$, where $n$ is the
total number of particles), so that every group is responsible for the
formation of its own excitations. In general, such a decomposition can
depend on time, while its character and the distribution of the groups in
the space of modes $N\left( k\right) $ constitute particular properties of
the medium which require the exact knowledge of the dynamics of particles
(in simple cases, however, this information can be found from simple
thermodynamical considerations, e.g., see Ref. \cite{K03}). As we will see,
the distribution of groups $N\left( k\right) $ in the space of modes defines
spectrum of thermal fluctuations in the system and, therefore, for practical
needs this function can be determined from direct measurements .

It is important that $q-$particles are identical, and, therefore, the
decomposition $n=\sum n_{i}$ has a conditional character. In particular, the
mean values $\rho _{0}$ and $u_{0}$ are determined by all groups, while
perturbations acquire analogous decomposition $A=\sum A_{i}$. The identity
of particles $q$ results in the analogous identity of fields $A_{i}$. Thus,
from the mathematical standpoint excitations will be particles which obey
the parastatistics, while the number of fields (groups) $N$ plays the role
of the order of the parastatistics.

In what follows we, for the sake of simplicity, neglect the presence of the
index $\alpha $ (which numerates different polarizations). In this manner,
when the number of fields is variable, the set of creation /annihilation
operators $\{a_{k},a_{k}^{+}\}$ is replaced by the expanded set $%
\{a_{k}\left( j\right) ,a_{k}^{+}\left( j\right) \}$, where $j\in \left[
1,...N\left( k\right) \right] $, and $N\left( k\right) $ is the total number
of fields (groups) for a given wave number $k$. Then the energy for a free
field can be written as

\begin{equation}
H_{0}=\sum_{k}\sum_{j=1}^{N\left( k\right) }\omega _{k}a_{k}^{+}\left(
j\right) a_{k}\left( j\right) \,.  \label{en}
\end{equation}
The fields are identical and are supposed to obey the same statistics as
that of $q-$particles. It is convenient to consider the set of
creation/annihilation operators for field modes $\left\{ C^{+}\left(
n,k\right) ,C\left( n,k\right) \right\} $, where $k$ is the wave number and $%
n$ is the number of particles in the given mode. Since the number of modes
corresponds to the true degrees of freedom there exists a relation between
the operators $C^{+}\left( n,k\right) $ and $C\left( n,k\right) $ and the
standard creation/annihilation operators $\psi ^{+}\left( x\right) $ and $%
\psi \left( x\right) $ for $q-$particles, i.e., $C\left( n,k\right) =\int
A\left( n,k,x\right) \psi \left( x\right) d^{3}x$ with some kernel $A\left(
n,k,x\right) $. Thus, operators $C^{+}\left( n,k\right) $ and $C\left(
n,k\right) $ obey the relations 
\begin{equation}
C\left( n,k\right) C^{+}\left( m,k^{\prime }\right) \pm C^{+}\left(
m,k^{\prime }\right) C\left( n,k\right) =\delta _{nm}\delta _{kk^{\prime
}}\;,  \label{c}
\end{equation}
where the sign reflects the statistics of $q-$particles. We note that in the
case of the so-called parafield theory there exists a spin-statistics
theorem \cite{gov} (analogous to the Pauli theorem) which states that
bosonic modes should be quantized according to the Fermi statistics, while
fermionic modes should obey the Bose statistics. This means that the fractal
behaviour can be associated only with fermions. In the case of a fractal
medium in which atoms have an integral spin we should assume that the
distribution of modes is determined by electrons. Mathematically, this can
be expressed by the fact that in the case when $q-$particles are bosons the
necessary kernel $A\left( n,k,x\right) $ does not exist. Thus, in what
follows we consider the case of the Fermi statistics for modes only (the
sign $+$ in (\ref{c})).

The eigenvalues of the Hamiltonian (\ref{en}) can be written
straightforwardly 
\begin{equation}
H_{0}=\sum_{k,n}\omega _{k}nN\left( n,k\right) ,
\end{equation}
where $N\left( n,k\right) =C^{+}\left( n,k\right) C\left( n,k\right) $ is
the number of field modes with the given wave number $k$ and the number of
phonons $n$. The fact that modes obey the identity principle results in a
redefinition of the standard creation and annihilation operators for phonons
(see, for details, Ref. \cite{K99}). Indeed, since fields $A_{i}$
(corresponding to different groups of $q-$particles) are identical they
cannot be detected separately. What is really measured in experiments is the
total perturbation $A=\sum A_{i}$ which is given by the expression (\ref{a})
in which, however, we should replace the creation/annihilation operators $%
a_{k}$ and $a_{k}^{+}$ by sums 
\begin{equation}
a_{k}=\sum_{j=1}^{N\left( k\right) }a_{k}\left( j\right)
,\,\;a_{k}^{+}=\sum_{j=1}^{N\left( k\right) }a_{k}^{+}\left( j\right) .
\end{equation}
In the representation of occupation numbers $N\left( n,k\right) $ these
operators have the expression \cite{K99} 
\begin{equation}
a_{k}=\sum_{n}\sqrt{n+1}C^{+}\left( n,k\right) C\left( n+1,k\right) ,\;
\label{op}
\end{equation}
\[
a_{k}^{+}=\sum_{n}\sqrt{n+1}C^{+}\left( n+1,k\right) C\left( n,k\right) \;. 
\]
Thus, from (\ref{c}) we find that these operators obey the relations 
\begin{equation}
a_{k}a_{k^{\prime }}^{+}-a_{k^{\prime }}^{+}a_{k}=N\left( k\right) \delta
_{k,k^{\prime }},  \label{crn}
\end{equation}
which generalize the standard relations (\ref{cr}) and are a consequence of
the fact that phonons obey the parastatistics. The order of the
parastatistics $N\left( k\right) $ is the operator of the total number of
field modes for a given wave number $k$ which is given by 
\begin{equation}
N\left( k\right) =\sum_{n}N\left( n,k\right) .
\end{equation}
Physically, this operator characterizes the distribution of the number of
groups of $q-$particles in the space of modes, while $N=\sum_{k}N\left(
k\right) $ gives the total number of particles in the system.

In the thermal equilibrium state (and in the absence of self-organization
processes) the number of fields is conserved and the operator $N\left(
k\right) $ can be considered as an ordinary function of wave numbers. Thus,
we assume that $N\left( k\right) $ represents a structural function of the
system which should remain constant (at least, for not very high
temperatures). Then the thermodynamically equilibrium state will be
characterized by mean values for occupation numbers of the type 
\begin{equation}
\left\langle N\left( k,n\right) \right\rangle =\left( \exp \left( \frac{%
n\omega _{k}-\mu _{k}}{T}\right) +1\right) ^{-1},  \label{on}
\end{equation}
where the chemical potential $\mu _{k}$ can be expressed via the structural
function $N\left( k\right) $ as 
\begin{equation}
N\left( k\right) =\sum_{n}\left( \exp \left( \frac{n\omega _{k}-\mu _{k}}{T}%
\right) +1\right) ^{-1}.  \label{N}
\end{equation}
In the limit $T\rightarrow 0$ the system reaches the ground state. From (\ref
{on}) we find that the ground state is characterized by the occupation
numbers 
\begin{equation}
N\left( n,k\right) =\theta \left( \mu _{k}-n\omega _{k}\right)  \label{gs}
\end{equation}
where $\theta \left( x\right) $ is the Heaviside step function. Thus, in the
limit $T\rightarrow 0$ we find that the chemical potential $\mu _{k}$ is
expressed via the structural function $N\left( k\right) $ as 
\begin{equation}
N\left( k\right) =\sum_{n}\theta \left( \mu _{k}-n\omega _{k}\right) =1+ 
\left[ \frac{\mu _{k}}{\omega _{k}}\right]  \label{gn}
\end{equation}
where $\left[ x\right] $ denotes the integral part of the number $x$.

From (\ref{gs}) we find that the ground state is characterized by a
nonvanishing number of phonons $n_{0}\left( k\right) $ 
\begin{equation}
n_{0}\left( k\right) =\sum_{n=0}^{\infty }nN\left( n,k\right) =\frac{1}{2}%
N\left( k\right) \left( N\left( k\right) -1\right)
\end{equation}
and corresponds to the energy $E_{0}=\sum \omega _{k}n_{0}\left( k\right) $.
This accounts for the fact that in the limit $T\rightarrow 0$ the
distribution of $q-$particles in the medium has an inhomogeneous (fractal)
character (i.e., in the ground state the density of $q-$particles contain a
specific dependence on coordinates $\rho _{0}\left( x\right) $, e.g. see
Ref. \cite{K03}).

The expression (\ref{gn}) allows to understand some general properties of
the structural function $N\left( k\right) $. Indeed, consider the case when $%
\mu _{k}=\mu $. Then there exists a region of wave numbers $\omega _{k}>$ $%
\mu $ in which $N\left( k\right) =1$ and phonons behave as standard Bose
particles. Thus, the wave number $k^{\ast }$ (at which $\omega =$ $\mu $)
characterizes the scale on which effects of the parastatistics start to show
up. In this sense regular systems can be considered as those systems in
which this scale $\ell ^{\ast }=2\pi /k^{\ast }$ exceeds the maximal
possible wavelength. At scales $k\ll k^{\ast }$ (i.e., as $\omega
_{k}\rightarrow 0$) we get $N\left( k\right) \sim \mu /\omega _{k}$ which
gives the fractal dimension $D\approx 2$.

Now we examine the problem how the structural function $N\left( k\right) $
can be expressed via the correlation functions and hence how it can be
directly measured. Substituting the expressions (\ref{op}) for the
creation/annihilation operators in (\ref{a}) we find that correlation
functions have the structure 
\begin{equation}
\left\langle A\left( x\right) ,A\left( x+r\right) \right\rangle =\frac{1}{%
2\pi ^{2}}\int_{0}^{\infty }\Phi ^{2}\left( k\right) \frac{\sin kr}{kr}\frac{%
dk}{k}
\end{equation}
where 
\begin{equation}
\Phi ^{2}\left( k\right) =k^{3}\left| e\left( k\right) \right| ^{2}\left( 2%
\widetilde{n}_{k}+N\left( k\right) \right)  \label{cf}
\end{equation}
and $\widetilde{n}_{k}=\left\langle a_{k}^{+}a_{k}\right\rangle $ is given
by 
\begin{equation}
\widetilde{n}_{k}=\sum_{n}\left( n+1\right) \left\langle N\left(
n+1,k\right) \right\rangle \left( 1-\left\langle N\left( n,k\right)
\right\rangle \right) ,
\end{equation}
where the occupation numbers $\left\langle N\left( n,k\right) \right\rangle $
are defined in (\ref{on}). For small temperatures $T\rightarrow 0$ we get $%
\widetilde{n}_{k}\rightarrow 0$ and the correlation function $\Phi
^{2}\left( k\right) $ explicitly defines the structural function of the
system $N\left( k\right) $.

The explicit form of the function $\widetilde{n}_{k}$ can be found as
follows. We recall that we assume that the number of modes $N\left( k\right) 
$ represents a constant function (the structural function of the system).
Then from (\ref{crn}) we see that the difference between operators $%
a_{k}^{+} $ and $a_{k}$ and the standard creation and annihilation operators
for Bose particles comes from the coefficient $a_{k}=\sqrt{N\left( k\right) }%
c_{k}$ (where $c_{k}$ is the standard annihilation operator, i.e., $\left[
c_{k},c_{k^{\prime }}^{+}\right] =\delta _{k,k^{\prime }}$). Hence for
thermal equilibrium state we should get the expression $\left\langle
a_{k}^{+}a_{k}\right\rangle =N\left( k\right) \left\langle
c_{k}^{+}c_{k}\right\rangle =N\left( k\right) \left( \exp \left( \frac{%
\omega }{T}\right) -1\right) ^{-1}$. Therefore, from the formal standpoint
the fact that phonons obey the parastatistics can be accounted for by the
substitution $\left( 2\overline{n}_{k}+1\right) $ $=$ $\coth \left( \frac{%
\omega }{2T}\right) $ $\rightarrow $ $N\left( k\right) \coth \left( \frac{%
\omega }{2T}\right) $ in all thermodynamical formulas and, in general, the
structural function $N\left( k\right) $ defines deviations of
thermodynamical potentials and fluctuations from the standard
thermodynamical laws. In particular, for phonons we get $\omega =ku$, where $%
u$ is the sound speed (in general $u$ can depend on the polarization of
phonons, but we neglect here this effect). Thus, if in some range of scales
a system possesses a fractal behaviour with dimension $D$, we find that in
that range fluctuations will acquire an amplification by the factor $N\left(
\omega \right) \sim \omega ^{-\alpha }$ with $\alpha =3-D$. Such kind of
amplification is indeed observed in various systems and represents the
so-called $1/f$ -noise problem \cite{Bu}. Hence we may conclude that at list
in some class of systems the origin of $1/f\,-$noise can have
straightforward relation to the fractal behaviour and, therefore, to the
parastatistics of elementary excitations. This problem, however, requires
the further investigation.

In this manner, we have shown that in a wide class of systems (which
demonstrate an anomalous behavior for thermodynamical potentials and
fluctuations, presumably at low frequencies) elementary excitations obey the
so-called parastatistics of a variable order $N\left( k\right) $. The order
of the parastatistics $N\left( k\right) $ reflects properties of a system at
low temperatures (e.g., the inhomogeneous distribution of particles) and
represents a specific, for every given system, structural function which can
be determined by means of direct measurements of fluctuations. In fractal
media this function behaves as $N\left( k\right) \sim k^{D-3}$, while for a
more general case (i.e., in general disordered systems) this function can
have an arbitrary behaviour.




\end{document}